\documentstyle[prb,aps]{revtex}

\begin{document}

\title{Parent-bilinear Sea-boson Correspondence with Spin}
\author{Girish S. Setlur and Yia-Chung Chang}
\address{Department of Physics and Materials Research Laboratory,\\
 University of Illinois at Urbana-Champaign , Urbana Il 61801}
\maketitle

\begin{abstract}
 This is a sort of tutorial where we present many of the tedious details
 involved in deriving the correspondence between the number conserving
 product of two fermi/bose fields with spin and the corresponding
 sea/condensate displacements also with spin. This time we include finite
 temperature effects differently by coupling the sea-bosons to an inhomogeneous
 chemical potential. It is shown that many of the salient features of the free
 theory are reproduced properly by this approach. An attempt is made to
 be as detailed as possible, this includes providing formulas for the
 sea-bosons as well as computing the finite temperature Fermi correlators.
\end{abstract}

\section{Parent Bilinear Sea-Boson Correspondence}

Without further ado(for more details the reader is refered to 
Ref.(\onlinecite{Setlur}))let us write down the correspondence,
\newline
\newline
{\Large{  For Fermi systems  }} :
\newline
\newline
\begin{equation}
[a_{ {\bf{k}}, \sigma }({\bf{q}}, \rho),
a_{ {\bf{k}}^{'}, \sigma^{'} }({\bf{q}}^{'}, \rho^{'})] = 0
\end{equation}
and,
\begin{equation}
[a_{ {\bf{k}}, \sigma }({\bf{q}}, \rho),
a^{\dagger}_{ {\bf{k}}^{'}, \sigma^{'} }({\bf{q}}^{'}, \rho^{'})] =
\delta_{ {\bf{k}}, {\bf{k}}^{'} }\delta_{ {\bf{q}}, {\bf{q}}^{'} }
 \delta_{ \sigma, \sigma^{'} }\delta_{ \rho, \rho^{'} }
\end{equation}
When $ {\bf{q}} \neq {\bf{0}} $,
\[
c^{\dagger}_{ {\bf{k}} + {\bf{q}}/2, \sigma }
c_{ {\bf{k}} - {\bf{q}}/2, \sigma^{'} }
 = \Lambda_{ {\bf{k}}, \sigma }({\bf{q}}, \sigma^{'})
a_{ {\bf{k}}, \sigma }(-{\bf{q}}, \sigma^{'})
+ a^{\dagger}_{ {\bf{k}}, \sigma^{'} }({\bf{q}}, \sigma)
\Lambda_{ {\bf{k}}, \sigma^{'} }(-{\bf{q}}, \sigma)
\]
\[
+ \sqrt{1 - {\bar{n}}_{ {\bf{k+q/2}}\sigma }}
 \sqrt{1 - {\bar{n}}_{ {\bf{k-q/2}}\sigma^{'} }}
       \sum_{ {\bf{q}}_{1}\sigma_{1} }
a^{\dagger}_{ {\bf{k+q/2-q_{1}/2}}\sigma_{1} }({\bf{q_{1}}}\sigma)
a_{ {\bf{k-q_{1}/2}}\sigma_{1} }({\bf{q_{1}-q}}\sigma^{'})
\]
\begin{equation}
- \sqrt{ {\bar{n}}_{ {\bf{k+q/2}}\sigma }}
\sqrt{{\bar{n}}_{ {\bf{k-q/2}}\sigma^{'} }}
 \sum_{ {\bf{q}}_{1}\sigma_{1} }
 a^{\dagger}_{ {\bf{k-q/2+q_{1}/2}}\sigma^{'} }({\bf{q_{1}}}\sigma_{1})
a_{ {\bf{k+q_{1}/2}}\sigma }({\bf{q_{1}-q}}\sigma_{1})
\end{equation}
here,
\begin{equation}
 \Lambda_{ {\bf{k}}, \sigma }({\bf{q}}, \sigma^{'}) =
\sqrt{
 {\bar{n}}_{ {\bf{k}} + {\bf{q}}/2, \sigma }
(1 - {\bar{n}}_{ {\bf{k}} - {\bf{q}}/2, \sigma^{'} })}
\end{equation}
Further,
\[
c^{\dagger}_{ {\bf{k}}, \sigma }c_{ {\bf{k}}, \sigma }
 = n_{ {\bf{k}}, \sigma } = n_{\beta}({\bf{k}})\frac{ N }
{\langle N \rangle}
+ \sum_{ {\bf{q}}, \sigma_{1} }
a^{\dagger}_{ {\bf{k}} - {\bf{q}}/2,\sigma_{1} }({\bf{q}},\sigma)
a_{ {\bf{k}} - {\bf{q}}/2,\sigma_{1} }({\bf{q}},\sigma)
\]
\begin{equation}
- \sum_{ {\bf{q}}, \sigma_{1} }
a^{\dagger}_{ {\bf{k}} + {\bf{q}}/2,\sigma }({\bf{q}},\sigma_{1})
a_{ {\bf{k}} + {\bf{q}}/2,\sigma }({\bf{q}},\sigma_{1})
\label{NUMBER}
\end{equation}
 We seldom have the need to deal with the object 
 $ c^{\dagger}_{ {\bf{k}} \sigma }c_{ {\bf{k}} \sigma^{'} } $ when
 $ \sigma \neq \sigma^{'} $, therefore we shall not write down a formula
 for this in terms of the sea-bosons.
Here,
\begin{equation}
 n_{\beta}({\bf{k}}) = \frac{1}{ exp(\beta(\epsilon_{ {\bf{k}} }-\mu)) + 1 }
\end{equation}
and,
\begin{equation}
 \sum_{ {\bf{k}}, \sigma }n_{\beta}({\bf{k}}) =  \langle N \rangle
\end{equation}
And the definition of the sea-boson is,
\begin{equation}
a_{ {\bf{k}} \sigma }({\bf{q}} \sigma^{'})
  = \frac{1}{\sqrt{ n_{ {\bf{k}}-{\bf{q}}/2 \sigma } }}
c^{\dagger}_{ {\bf{k}}-{\bf{q}}/2 \sigma }
(\frac{ n^{\beta}({\bf{k}}-{\bf{q}}/2) }
{\langle N \rangle })
^{\frac{1}{2}}
e^{i\mbox{   }\theta({\bf{k}},{\bf{q}};\sigma,\sigma^{'})}
c_{ {\bf{k}}+{\bf{q}}/2 \sigma^{'} }
\label{DEFN}
\end{equation}
In order to prove this we have to verify the following facts.
\newline
\newline
(a) The definition of the sea-boson in Eq.(~\ref{DEFN}) when plugged into
 the formula for the number operator in  Eq.(~\ref{NUMBER}) gives an identity.
\newline
\newline
(b) The dynamical four-point and six-point functions are recovered
 correctly.
\newline
\newline
(c) The commutation rule between the number operator and the RPA
 form of the off-diagonal fermi bilinear is correctly reproduced.
\newline
\newline
(d) The commutation rule between two off-diagonal fermi bilinears
 is reproduced in the RPA sense.
\newline
\newline
Let us verify assertion (a).
\[
\sum_{ {\bf{q}} \sigma_{1} }
a^{\dagger}_{ {\bf{k}}-{\bf{q}}/2\sigma_{1} }({\bf{q}}\sigma)
a_{ {\bf{k}}-{\bf{q}}/2\sigma_{1} }({\bf{q}}\sigma)
 = \sum_{ {\bf{q}} \sigma_{1} }
\frac{n^{\beta}({\bf{k}}-{\bf{q}})n_{ {\bf{k}}\sigma } }{\langle N \rangle }
 = n_{ {\bf{k}}\sigma }
\]
\[
\sum_{ {\bf{q}} \sigma_{1} }
a^{\dagger}_{ {\bf{k}}+{\bf{q}}/2\sigma }({\bf{q}}\sigma_{1})
a_{ {\bf{k}}+{\bf{q}}/2\sigma }({\bf{q}}\sigma_{1})
 = \sum_{ {\bf{q}} \sigma_{1} }
\frac{n^{\beta}({\bf{k}})n_{ {\bf{k-q}}\sigma_{1} } }{\langle N \rangle }
 = n^{\beta}({\bf{k}})\frac{ N }{ \langle N \rangle }
\]
and thus Eq.(~\ref{NUMBER}) follows.
As far as assertion (b) is concerned,
let us first check to see if the time-evolved versions of the
off-diagonal fermi-bilinear is right.
The kinetic energy operator is,
\begin{equation}
K = \sum_{ {\bf{k}}, {\bf{q}}, \sigma, \sigma_{1} }
(\frac{ {\bf{k.q}} }{m})a^{\dagger}_{ {\bf{k}} \sigma }({\bf{q}}\sigma_{1})
a_{ {\bf{k}} \sigma }({\bf{q}}\sigma_{1})
\end{equation}
From this it is easy to see that the time-evolved version is
\[
c^{\dagger}_{ {\bf{k}} + {\bf{q}}/2 \sigma }
c_{ {\bf{k}} - {\bf{q}}/2 \sigma^{'} }(t)
 = c^{\dagger}_{ {\bf{k}} + {\bf{q}}/2 \sigma }
c_{ {\bf{k}} - {\bf{q}}/2 \sigma^{'} }(t=0)
e^{it(\frac{ {\bf{k.q}} }{m})}
\]
The four-point function is given by,
\[
\langle c^{\dagger}_{ {\bf{k}}_{a} + {\bf{q}}_{a}/2 \sigma_{a} }
c_{ {\bf{k}}_{a} - {\bf{q}}_{a}/2 \sigma_{a}^{'} }
c^{\dagger}_{ {\bf{k}}_{b} + {\bf{q}}_{b}/2 \sigma_{b} }
c_{ {\bf{k}}_{b} - {\bf{q}}_{b}/2 \sigma_{b}^{'} }
 \rangle
\]
\[
 =  n_{ {\bf{k}}_{a} + {\bf{q}}_{a}/2 \sigma_{a} }
(1 - n_{ {\bf{k}}_{a} - {\bf{q}}_{a}/2 \sigma_{a}^{'} })
\delta_{ {\bf{k}}_{a} + {\bf{q}}_{a}/2, {\bf{k}}_{b} - {\bf{q}}_{b}/2 }
\delta_{ \sigma_{a}, \sigma_{b}^{'} }
\delta_{ {\bf{k}}_{a} - {\bf{q}}_{a}/2, {\bf{k}}_{b} + {\bf{q}}_{b}/2 }
\delta_{ \sigma_{a}^{'}, \sigma_{b} }
\]
\[
= \Lambda^{2}_{ {\bf{k}}_{a} \sigma_{a} }({\bf{q}}_{a}\sigma^{'}_{a})
\delta_{ {\bf{k}}_{a}, {\bf{k}}_{b} }
\delta_{ {\bf{q}}_{a}, {\bf{q}}_{b} }
\delta_{ \sigma_{a}, \sigma_{b}^{'} }\delta_{ \sigma_{a}^{'}, \sigma_{b} }
\]
In order to verify the six-point function, we proceed as follows.
From elementary considerations it is clear that,
\[
\langle c^{\dagger}_{ {\bf{k}}_{a} + {\bf{q}}_{a}/2 \sigma_{a} }
c_{ {\bf{k}}_{a} - {\bf{q}}_{a}/2 \sigma_{a}^{'} }
c^{\dagger}_{ {\bf{k}}_{b} + {\bf{q}}_{b}/2 \sigma_{b} }
c_{ {\bf{k}}_{b} - {\bf{q}}_{b}/2 \sigma_{b}^{'} }
c^{\dagger}_{ {\bf{k}}_{c} + {\bf{q}}_{c}/2 \sigma_{c} }
c_{ {\bf{k}}_{c} - {\bf{q}}_{c}/2 \sigma_{c}^{'} }
\rangle
\]
\[
 = \delta_{ {\bf{k}}_{a} + {\bf{q}}_{a}/2, {\bf{k}}_{c} - {\bf{q}}_{c}/2 }
\delta_{ \sigma_{a}, \sigma_{c}^{'} }
\delta_{ {\bf{k}}_{a} - {\bf{q}}_{a}/2, {\bf{k}}_{b} + {\bf{q}}_{b}/2 }
 \delta_{  \sigma_{a}^{'}, \sigma_{b} }
\delta_{ {\bf{k}}_{b} - {\bf{q}}_{b}/2, {\bf{k}}_{c} + {\bf{q}}_{c}/2 }
\delta_{ \sigma_{b}^{'}, \sigma_{c} }
\]
\[
\times
n_{ {\bf{k}}_{a} + {\bf{q}}_{a}/2 \sigma_{a} }
(1-n_{ {\bf{k}}_{a} - {\bf{q}}_{a}/2 \sigma_{a}^{'} })
(1-n_{{\bf{k}}_{b} - {\bf{q}}_{b}/2 \sigma_{b}^{'} })
\]
\begin{equation}
 - \delta_{ {\bf{k}}_{a} + {\bf{q}}_{a}/2 , {\bf{k}}_{b} - {\bf{q}}_{b}/2}
 \delta_{ \sigma_{a}, \sigma_{b}^{'} }
\delta_{ {\bf{k}}_{a} - {\bf{q}}_{a}/2, {\bf{k}}_{c} + {\bf{q}}_{c}/2 }
\delta_{ \sigma_{a}^{'}, \sigma_{c} }
\delta_{ {\bf{k}}_{b} + {\bf{q}}_{b}/2, {\bf{k}}_{c} - {\bf{q}}_{c}/2 }
\delta_{ \sigma_{b}, \sigma_{c}^{'} }
n_{ {\bf{k}}_{a} + {\bf{q}}_{a}/2 \sigma_{a} }
n_{ {\bf{k}}_{b} + {\bf{q}}_{b}/2 \sigma_{b} }
(1 - n_{ {\bf{k}}_{a} - {\bf{q}}_{a}/2 \sigma_{a}^{'} })
\label{A}
\end{equation}
In the sea-boson language the same quantity is,
\[
\langle \Lambda_{ {\bf{k}}_{a} \sigma_{a} }({\bf{q}}_{a}\sigma_{a}^{'})
a_{ {\bf{k}}_{a} \sigma_{a} }(-{\bf{q}}_{a}\sigma_{a}^{'})
\sqrt{1 - n_{ {\bf{k}}_{b} + {\bf{q}}_{b}/2\sigma_{b} }}
\sqrt{1 - n_{ {\bf{k}}_{b} - {\bf{q}}_{b}/2\sigma_{b}^{'} }}
\]
\[
\times
\sum_{ {\bf{q}}_{1} \sigma_{1} }
a^{\dagger}_{ {\bf{k}}_{b}+{\bf{q}}_{b}/2
- {\bf{q}}_{1}/2 \sigma_{1} }({\bf{q}}_{1}\sigma_{b})
a_{ {\bf{k}}_{b}- {\bf{q}}_{1}/2 \sigma_{1} }
({\bf{q}}_{1}-{\bf{q}}_{b}\sigma_{b}^{'})
\Lambda_{ {\bf{k}}_{c} \sigma_{c}^{'} }(-{\bf{q}}_{c}\sigma_{c})
a^{\dagger}_{ {\bf{k}}_{c} \sigma_{c}^{'} }({\bf{q}}_{c}\sigma_{c})
\rangle
\]
\[
= \Lambda_{ {\bf{k}}_{a} \sigma_{a} }({\bf{q}}_{a}\sigma_{a}^{'})
\sqrt{1 - n_{ {\bf{k}}_{b} + {\bf{q}}_{b}/2\sigma_{b} }}
\sqrt{1 - n_{ {\bf{k}}_{b} - {\bf{q}}_{b}/2\sigma_{b}^{'} }}
\Lambda_{ {\bf{k}}_{c} \sigma_{c}^{'} }(-{\bf{q}}_{c}\sigma_{c})
\delta_{ {\bf{k}}_{a} = {\bf{k}}_{b}+{\bf{q}}_{b}/2+{\bf{q}}_{a}/2 }
\]
\[
\delta_{ \sigma_{a}, \sigma_{c}^{'} }
\delta_{ \sigma_{b}, \sigma_{a}^{'} }
\delta_{ {\bf{k}}_{c} = {\bf{k}}_{b}+{\bf{q}}_{a}/2 }
\delta_{ {\bf{q}}_{a}+ {\bf{q}}_{b} + {\bf{q}}_{c}, 0 }
\delta_{ \sigma_{c}, \sigma_{b}^{'} }
\]
\[
 = n_{ {\bf{k}}_{a}+{\bf{q}}_{a}/2 \sigma_{a} }
(1 - n_{ {\bf{k}}_{b} + {\bf{q}}_{b}/2\sigma_{b} })
(1 - n_{ {\bf{k}}_{b} - {\bf{q}}_{b}/2\sigma_{b}^{'} })
\]
\begin{equation}
\times
\delta_{ \sigma_{a}, \sigma_{c}^{'} }
\delta_{ \sigma_{c}, \sigma_{b}^{'} }
\delta_{ \sigma_{b}, \sigma_{a}^{'} }
\delta_{ {\bf{k}}_{a}-{\bf{q}}_{a}/2, {\bf{k}}_{b}+{\bf{q}}_{b}/2 }
\delta_{ {\bf{k}}_{b}-{\bf{q}}_{b}/2,  {\bf{k}}_{c}+{\bf{q}}_{c}/2 }
\delta_{ {\bf{k}}_{a}+{\bf{q}}_{a}/2, {\bf{k}}_{c}-{\bf{q}}_{c}/2 }
\label{EQN1}
\end{equation}
\[
-\langle \Lambda_{ {\bf{k}}_{a} \sigma_{a} }({\bf{q}}_{a}\sigma_{a}^{'})
a_{ {\bf{k}}_{a} \sigma_{a} }(-{\bf{q}}_{a}\sigma_{a}^{'})
\sqrt{n_{ {\bf{k}}_{b} + {\bf{q}}_{b}/2\sigma_{b} }}
\sqrt{n_{ {\bf{k}}_{b} - {\bf{q}}_{b}/2\sigma_{b}^{'} }}
\]
\[
\times
\sum_{ {\bf{q}}_{1} \sigma_{1} }
a^{\dagger}_{ {\bf{k}}_{b}-{\bf{q}}_{b}/2
+ {\bf{q}}_{1}/2 \sigma_{b}^{'} }({\bf{q}}_{1}\sigma_{1})
a_{ {\bf{k}}_{b} + {\bf{q}}_{1}/2 \sigma_{b} }
({\bf{q}}_{1}-{\bf{q}}_{b}\sigma_{1})
\Lambda_{ {\bf{k}}_{c} \sigma_{c}^{'} }(-{\bf{q}}_{c}\sigma_{c})
a^{\dagger}_{ {\bf{k}}_{c} \sigma_{c}^{'} }({\bf{q}}_{c}\sigma_{c})
\rangle
\]
\[
= -\Lambda_{ {\bf{k}}_{a} \sigma_{a} }({\bf{q}}_{a}\sigma_{a}^{'})
\sqrt{n_{ {\bf{k}}_{b} + {\bf{q}}_{b}/2\sigma_{b} }}
\sqrt{n_{ {\bf{k}}_{b} - {\bf{q}}_{b}/2\sigma_{b}^{'} }}
\Lambda_{ {\bf{k}}_{c} \sigma_{c}^{'} }(-{\bf{q}}_{c}\sigma_{c})
\]
\[
\times
\delta_{ {\bf{k}}_{a}, {\bf{k}}_{b}-{\bf{q}}_{b}/2 - {\bf{q}}_{a}/2 }
\delta_{ \sigma_{a}, \sigma_{b}^{'} }
\delta_{ \sigma_{a}^{'}, \sigma_{c} }
\delta_{ \sigma_{c}^{'}, \sigma_{b} }
\delta_{  {\bf{k}}_{b} - {\bf{q}}_{a}/2 , {\bf{k}}_{c} }
\delta_{ {\bf{q}}_{a} + {\bf{q}}_{b} + {\bf{q}}_{c}, 0 }
\]
\begin{equation}
= -n_{ {\bf{k}}_{b} + {\bf{q}}_{b}/2\sigma_{b} }
n_{ {\bf{k}}_{b} - {\bf{q}}_{b}/2\sigma_{b}^{'} }
(1-n_{ {\bf{k}}_{a} - {\bf{q}}_{a}/2\sigma_{a}^{'} })
\delta_{ {\bf{k}}_{a}, {\bf{k}}_{b}-{\bf{q}}_{b}/2 - {\bf{q}}_{a}/2 }
\delta_{ \sigma_{a}, \sigma_{b}^{'} }
\delta_{ \sigma_{a}^{'}, \sigma_{c} }
\delta_{ \sigma_{c}^{'}, \sigma_{b} }
\delta_{  {\bf{k}}_{b} - {\bf{q}}_{a}/2 , {\bf{k}}_{c} }
\delta_{ {\bf{q}}_{a} + {\bf{q}}_{b} + {\bf{q}}_{c}, 0 }
\label{EQN2}
\end{equation}
The sum of  Eq.(~\ref{EQN1}) and Eq.(~\ref{EQN2}) is the same as Eq.(~\ref{A}).
Therefore all is well as regards assertion (b). As far as (c) is concerned,
we merely compute the commutator:
\[
[c^{\dagger}_{ {\bf{k}}+{\bf{q}}/2\sigma }c_{ {\bf{k}}-{\bf{q}}/2\sigma^{'} },
n_{ {\bf{p}} \lambda }]
 = [(\Lambda_{ {\bf{k}}\sigma }({\bf{q}}\sigma^{'})
a_{ {\bf{k}}\sigma }(-{\bf{q}}\sigma^{'})
+ a^{\dagger}_{ {\bf{k}}\sigma^{'} }({\bf{q}}\sigma)
\Lambda_{ {\bf{k}}\sigma^{'} }(-{\bf{q}}\sigma)),
\sum_{ {\bf{q}}_{1} \sigma_{1} }
a^{\dagger}_{ {\bf{p}}-{\bf{q}}_{1}/2\sigma_{1} }({\bf{q}}_{1}\lambda)
a_{ {\bf{p}}-{\bf{q}}_{1}/2\sigma_{1} }({\bf{q}}_{1}\lambda)]
\]
\[
- [(\Lambda_{ {\bf{k}}\sigma }({\bf{q}}\sigma^{'})
a_{ {\bf{k}}\sigma }(-{\bf{q}}\sigma^{'})
+ a^{\dagger}_{ {\bf{k}}\sigma^{'} }({\bf{q}}\sigma)
\Lambda_{ {\bf{k}}\sigma^{'} }(-{\bf{q}}\sigma)),
\sum_{ {\bf{q}}_{1} \sigma_{1} }
a^{\dagger}_{ {\bf{p}}+{\bf{q}}_{1}/2\lambda }({\bf{q}}_{1}\sigma_{1})
a_{ {\bf{p}}+{\bf{q}}_{1}/2\lambda }({\bf{q}}_{1}\sigma_{1})]
\]
\[
= \Lambda_{ {\bf{k}}\sigma }({\bf{q}}\sigma^{'})
\delta_{ {\bf{k}}, {\bf{p}}+{\bf{q}}/2 }\delta_{ \lambda,\sigma^{'} }
a_{ {\bf{k}}\sigma }(-{\bf{q}}\sigma^{'})
- \Lambda_{ {\bf{k}}\sigma^{'} }(-{\bf{q}}\sigma)
\delta_{ {\bf{k}}, {\bf{p}}-{\bf{q}}/2 }
\delta_{ \lambda,\sigma }
a^{\dagger}_{ {\bf{k}}\sigma^{'} }({\bf{q}}\sigma)
\]
\[
-\Lambda_{ {\bf{k}}\sigma }({\bf{q}}\sigma^{'})
\delta_{ {\bf{k}}, {\bf{p}}-{\bf{q}}/2 }\delta_{ \lambda,\sigma }
a_{ {\bf{k}}\sigma }(-{\bf{q}}\sigma^{'})
+ \Lambda_{ {\bf{k}}\sigma^{'} }(-{\bf{q}}\sigma)
\delta_{ {\bf{k}}, {\bf{p}}+{\bf{q}}/2 }\delta_{ \lambda,\sigma^{'} }
a^{\dagger}_{ {\bf{k}}\sigma^{'} }({\bf{q}}\sigma)
\]
\[
= \delta_{ {\bf{k}}, {\bf{p}}+{\bf{q}}/2 }\delta_{ \lambda, \sigma^{'} }
c^{\dagger}_{ {\bf{k}}+{\bf{q}}/2\sigma}c_{ {\bf{k}}-{\bf{q}}/2\sigma^{'}} 
-\delta_{ {\bf{k}}, {\bf{p}}-{\bf{q}}/2 }\delta_{ \lambda, \sigma }
c^{\dagger}_{ {\bf{k}}+{\bf{q}}/2\sigma}c_{ {\bf{k}}-{\bf{q}}/2\sigma^{'}}
\]
 which is as it should be.
 The commutation rule between the off-diagonal fermi bilinears is given by,
\[
[c^{\dagger}_{ {\bf{k}}+{\bf{q}}/2\sigma }c_{ {\bf{k}}-{\bf{q}}/2\rho },
c^{\dagger}_{ {\bf{k}}^{'}+{\bf{q}}^{'}/2\sigma^{'} }
c_{ {\bf{k}}^{'}-{\bf{q}}^{'}/2\rho^{'} }]
 = [(\Lambda_{ {\bf{k}} \sigma }({\bf{q}}\rho)
a_{ {\bf{k}} \sigma }(-{\bf{q}}\rho)
+ \Lambda_{ {\bf{k}} \rho }(-{\bf{q}}\sigma)
a^{\dagger}_{ {\bf{k}} \rho }({\bf{q}}\sigma)),
\]
\[
(\Lambda_{ {\bf{k}}^{'} \sigma^{'} }({\bf{q}}^{'}\rho^{'})
a_{ {\bf{k}}^{'} \sigma^{'} }(-{\bf{q}}^{'}\rho^{'})
+ \Lambda_{ {\bf{k}}^{'} \rho^{'} }(-{\bf{q}}^{'}\sigma^{'})
a^{\dagger}_{ {\bf{k}}^{'} \rho^{'} }({\bf{q}}^{'}\sigma^{'}))]
\]
\begin{equation}
= ( {\bar{n}}_{ {\bf{k}}+{\bf{q}}/2 \sigma } - {\bar{n}}_{ {\bf{k}}-{\bf{q}}/2 
\rho } ) 
\delta_{ {\bf{k}}, {\bf{k}}^{'} }\delta_{ \sigma, \rho^{'} }
\delta_{ -{\bf{q}}, {\bf{q}}^{'} }\delta_{ \rho, \sigma^{'} }
\end{equation}
which is true only in the RPA-sense.

\subsection{Finite Temperature Aspects}

 In order to ensure that the proper finite temperature correlators are
 reproduced, it seems that we have  to couple the sea-bosons to an
 inhomogeneous chemical potential. It will be shown that this 
 prescription does the job it is required to do, namely it reproduces
 four-point and six point functions at finite temperature even when the
 sea-bosons are allowed to participate in the thermodynamic averaging.
 To do this we examine the object,
\begin{equation}
\langle a^{\dagger}_{ {\bf{k}} \sigma }({\bf{q}} \sigma^{'})
 a_{ {\bf{k}} \sigma }({\bf{q}} \sigma^{'})  \rangle
 = \frac{ n^{\beta}( {\bf{k}} + {\bf{q}}/2 )
n^{\beta}( {\bf{k}} - {\bf{q}}/2 ) }{ \langle N \rangle }
\end{equation}
 and in the sea-boson language it is,
\begin{equation}
\langle a^{\dagger}_{ {\bf{k}} \sigma }({\bf{q}} \sigma^{'})
 a_{ {\bf{k}} \sigma }({\bf{q}} \sigma^{'})  \rangle
 = \frac{1}{exp(\beta( \frac{ {\bf{k.q}} }{m} - \mu({\bf{k}},{\bf{q}}) ))
 - 1}
\end{equation}
therefore,
\begin{equation}
\mu({\bf{k}},{\bf{q}})
 =  \frac{ {\bf{k.q}} }{m} - (k_{B}T)Log(1 + \frac{ \langle N \rangle}
{ n^{\beta}( {\bf{k}} + {\bf{q}}/2 )
n^{\beta}( {\bf{k}} - {\bf{q}}/2 ) } )
\end{equation}
 The four-point and the six-point 
 functions may be evaluated correctly even though this
 time (as opposed to last time as in Ref(\onlinecite{Setlur}))
 the sea-bosons do participate in the thermodynamic averaging.
 In order to see this let us compute the four-point function,
\[
\langle c^{\dagger}_{ {\bf{k}}_{1}+{\bf{q}}_{1}/2 \sigma_{1} }
c_{ {\bf{k}}_{1}-{\bf{q}}_{1}/2 \sigma^{'}_{1} }
c^{\dagger}_{ {\bf{k}}_{2}-{\bf{q}}_{2}/2 \sigma^{'}_{2} }
c_{ {\bf{k}}_{2}+{\bf{q}}_{2}/2 \sigma_{2} }
\rangle
\]
\[
 = \Lambda_{ {\bf{k}}_{1} \sigma_{1} }({\bf{q}}_{1}\sigma^{'}_{1})
\Lambda_{ {\bf{k}}_{2} \sigma_{2} }({\bf{q}}_{2}\sigma^{'}_{2})
\langle a_{ {\bf{k}}_{1} \sigma_{1}  }(-{\bf{q}}_{1}\sigma^{'}_{1})
a^{\dagger}_{ {\bf{k}}_{2} \sigma_{2}  }(-{\bf{q}}_{2}\sigma^{'}_{2})
 \rangle
\]
\[
+ \Lambda_{ {\bf{k}}_{2} \sigma^{'}_{2}  }
(-{\bf{q}}_{2}\sigma_{2})
 \Lambda_{ {\bf{k}}_{2} \sigma^{'}_{2}  }
(-{\bf{q}}_{2}\sigma_{2})
\langle a^{\dagger}_{ {\bf{k}}_{1} \sigma^{'}_{1}  }
({\bf{q}}_{1}\sigma_{1})
a_{ {\bf{k}}_{2} \sigma^{'}_{2}  }({\bf{q}}_{2}\sigma_{2})
 \rangle
\]
\begin{equation}
 = \Lambda^{2}_{ {\bf{k}}_{1} \sigma_{1} }({\bf{q}}_{1}\sigma^{'}_{1})
\delta_{ {\bf{k}}_{1}, {\bf{k}}_{2} }
\delta_{ {\bf{q}}_{1}, {\bf{q}}_{2} }
\delta_{ \sigma_{1}, \sigma_{2} }
\delta_{ \sigma^{'}_{1}, \sigma^{'}_{2} }
 + O(\frac{1}{\langle N \rangle })
\end{equation}
Since the expectation value 
\begin{equation}
 \langle a_{ {\bf{k}}_{1} \sigma_{1}  }(-{\bf{q}}_{1}\sigma^{'}_{1})
a^{\dagger}_{ {\bf{k}}_{2} \sigma_{2}  }(-{\bf{q}}_{2}\sigma^{'}_{2})
 \rangle = \delta_{ {\bf{k}}_{1}, {\bf{k}}_{2} }
\delta_{ \sigma_{1}, \sigma_{2}  }
\delta_{ {\bf{q}}_{1}, {\bf{q}}_{2} }
\delta_{ \sigma^{'}_{1}, \sigma^{'}_{2} }
(1 +  O(\frac{1}{\langle N \rangle }))
\end{equation}
and,
\begin{equation}
\langle a^{\dagger}_{ {\bf{k}}_{1} \sigma^{'}_{1}  }
({\bf{q}}_{1}\sigma_{1})
a_{ {\bf{k}}_{2} \sigma^{'}_{2}  }({\bf{q}}_{2}\sigma_{2})
 \rangle
 =  \delta_{ {\bf{k}}_{1}, {\bf{k}}_{2} }
\delta_{ \sigma_{1}, \sigma_{2}  }
\delta_{ {\bf{q}}_{1}, {\bf{q}}_{2} }
\delta_{ \sigma^{'}_{1}, \sigma^{'}_{2} }O(\frac{1}{\langle N \rangle })
\end{equation}
and therefore all is well as regards thermodynamic expectation values.
(We assume here that we are dealing with a large system and  
 $ \langle N \rangle $ is large enough)
\newpage

{\Large{For Bose Systems}} : 
\newline
\newline
\newline
The definition of the condensate boson is as follows
( $ {\bf{q}} \neq {\bf{0}} $ ),
\begin{equation}
d_{ \sigma }({\bf{q}}\sigma^{'})
 = \frac{1}{ \sqrt{N_{0\sigma}} }b^{\dagger}_{ {\bf{0}}\sigma}
b_{ {\bf{q}}\sigma^{'} }
\label{DEFN2}
\end{equation}
however,
\begin{equation}
d_{ \sigma }({\bf{0}}\sigma^{'}) = 0
\end{equation}
\begin{equation}
[d_{ \sigma }({\bf{q}}\rho), 
d_{ \sigma^{'} }({\bf{q}}^{'}\rho^{'})] = 0
\end{equation}

\begin{equation}
[d_{ \sigma }({\bf{q}}\rho),
d^{\dagger}_{ \sigma^{'} }({\bf{q}}^{'}\rho^{'})] = 
e^{i(X_{ {\bf{0}}\sigma }-X_{ {\bf{0}}\sigma^{'} })}
\delta_{ {\bf{q}}, {\bf{q}}^{'} }
\delta_{ \rho, \rho^{'} }
\end{equation}
where $ [X_{ {\bf{0}}\sigma },N_{ {\bf{0}}\sigma^{'} }] = i\delta_{ \sigma, \sigma^{'} } $
however, we shall be lax in most cases and replace 
$ N_{ {\bf{0}}\sigma^{'} } $ by
 $ \langle N_{ {\bf{0}}\sigma^{'} } \rangle $, in this case
 we have to assume that $ X_{ {\bf{0}}\sigma } $ is actually infinite
 in order that it is the canonical conjugate of the c-number
 $ \langle N_{ {\bf{0}}\sigma } \rangle $.
In this case the condensate bosons are indeed exact bosons obeying,
\begin{equation}
[d_{ \sigma }({\bf{q}}\rho),
d^{\dagger}_{ \sigma^{'} }({\bf{q}}^{'}\rho^{'})] =
\delta_{  \sigma, \sigma^{'} }
\delta_{ {\bf{q}}^{'}, {\bf{q}} }\delta_{ \rho, \rho^{'} }
\end{equation}
when $ {\bf{q}} \neq 0 $,
\[
b^{\dagger}_{ {\bf{k}}+{\bf{q}}/2 \sigma }
b_{ {\bf{k}}-{\bf{q}}/2 \sigma^{'} }
 = \sqrt{ N_{0\sigma} }d_{ \sigma }(-{\bf{q}}\sigma^{'})
\delta_{ {\bf{k}}+{\bf{q}}/2, {\bf{0}} }
+ d^{\dagger}_{ \sigma^{'} }({\bf{q}}\sigma)
 \sqrt{ N_{0\sigma^{'}} }
\delta_{ {\bf{k}}-{\bf{q}}/2, {\bf{0}} }
\]
\begin{equation}
+ d^{\dagger}_{ \sigma_{1} }
({\bf{k}}+{\bf{q}}/2\sigma)
d_{ \sigma_{1} }({\bf{k}}-{\bf{q}}/2\sigma^{'})
\label{OFFDCORR}
\end{equation}
where $ \sigma_{1} $ is anything.
\begin{equation}
b^{\dagger}_{ {\bf{k}}\sigma }b_{ {\bf{k}}\sigma }
 = N_{0\sigma }\delta_{ {\bf{k}}, {\bf{0}} }
 + d^{\dagger}_{ \sigma }({\bf{k}}\sigma)
d_{ \sigma }({\bf{k}}\sigma)
\label{DCORR}
\end{equation}
and,
\begin{equation}
N_{0\sigma } = N_{ \sigma } - 
\sum_{ {\bf{q}} \neq 0 }
d^{\dagger}_{ \sigma }({\bf{q}}\sigma)
d_{ \sigma }({\bf{q}}\sigma)
\end{equation}
The above correspondence reproduces the salient
 features of the free theory such as,

(a) The definition of the condensate boson in Eq.(~\ref{DEFN2}) when
    plugged into the Eq.(~\ref{OFFDCORR}) and Eq.(~\ref{DCORR})
    give identities.

(b) The dynamical four-point and six-point functions of the
 free theory are reproduced properly.

(c) The condensate boson satisfies canonical Bose commutation rules
 and this fact is transparent(almost!) unlike in the Fermi case.

 Assertion (a) is obvious and we have already demonstrated (c), now we move
 on to (b). In order to verify (b), we have to compute the finite
 temperature four-point and six-point functions
 ($ {\bf{q}}_{1} \neq {\bf{0}} $$ {\bf{q}}_{2} \neq {\bf{0}} $),
\[
\langle b^{\dagger}_{ {\bf{k}}_{1}+{\bf{q}}_{1}/2\sigma_{1} }
b_{ {\bf{k}}_{1}-{\bf{q}}_{1}/2\sigma^{'}_{1} }
b^{\dagger}_{ {\bf{k}}_{2}-{\bf{q}}_{2}/2\sigma^{'}_{2} }
b_{ {\bf{k}}_{2}+{\bf{q}}_{2}/2\sigma_{2} }
\rangle
\]
\[
 =\delta_{ {\bf{k}}_{1}+{\bf{q}}_{1}/2, {\bf{0}} }
\delta_{  {\bf{k}}_{2}+{\bf{q}}_{2}/2 , {\bf{0}} }
 \sqrt{ \langle N_{ {\bf{0}} \sigma_{1} } \rangle }
 \sqrt{ \langle N_{ {\bf{0}} \sigma_{2} } \rangle }
\langle d_{ \sigma_{1} }(-{\bf{q}}_{1}\sigma^{'}_{1})
d^{\dagger}_{ \sigma_{2} }(-{\bf{q}}_{2}\sigma^{'}_{2}) \rangle
\]
\begin{equation}
+ \sqrt{ \langle N_{ {\bf{0}} \sigma^{'}_{1} } \rangle }
\sqrt{ \langle N_{ {\bf{0}} \sigma^{'}_{2} } \rangle }
\delta_{ {\bf{k}}_{1}-{\bf{q}}_{1}/2, {\bf{0}} }
\delta_{  {\bf{k}}_{2}-{\bf{q}}_{2}/2 , {\bf{0}} }
\langle d^{\dagger}_{  \sigma^{'}_{1} }({\bf{q}}_{1}\sigma_{1})
 d_{  \sigma^{'}_{2} }({\bf{q}}_{2}\sigma_{2}) \rangle
\end{equation}
Also,
\begin{equation}
\langle d^{\dagger}_{  \sigma^{'}_{1} }({\bf{q}}_{1}\sigma_{1})
 d_{  \sigma^{'}_{2} }({\bf{q}}_{2}\sigma_{2}) \rangle 
 = e^{i(X_{ {\bf{0}} \sigma_{2}^{'} }- X_{ {\bf{0}} \sigma_{1}^{'} })}
\langle 
 b^{\dagger}_{ {\bf{q}}_{1}\sigma_{1} }b_{ {\bf{q}}_{2}\sigma_{2} }
\rangle
 = 0
\end{equation}

\begin{equation}
\langle d_{ \sigma_{1} }(-{\bf{q}}_{1}\sigma^{'}_{1})
d^{\dagger}_{ \sigma_{2} }(-{\bf{q}}_{2}\sigma^{'}_{2}) \rangle
 = e^{i(X_{ {\bf{0}} \sigma_{1} }-X_{ {\bf{0}} \sigma_{2} })}
\langle b_{ -{\bf{q}}_{1}\sigma^{'}_{1} }
b^{\dagger}_{ -{\bf{q}}_{2}\sigma^{'}_{2} }\rangle
 = \delta_{ \sigma_{1} , \sigma_{2} }
\delta_{ {\bf{q}}_{1}, {\bf{q}}_{2} }
\delta_{ \sigma^{'}_{1}, \sigma^{'}_{2} }
\end{equation}
Therefore we have,
\begin{equation}
\langle b^{\dagger}_{ {\bf{k}}_{1}+{\bf{q}}_{1}/2\sigma_{1} }
b_{ {\bf{k}}_{1}-{\bf{q}}_{1}/2\sigma^{'}_{1} }
b^{\dagger}_{ {\bf{k}}_{2}-{\bf{q}}_{2}/2\sigma^{'}_{2} }
b_{ {\bf{k}}_{2}+{\bf{q}}_{2}/2\sigma_{2} }
\rangle
 =
 \langle N_{ {\bf{0}} \sigma_{1} } \rangle
\delta_{ {\bf{q}}_{1}, {\bf{q}}_{2} }
\delta_{ {\bf{k}}_{1}, {\bf{k}}_{2} }
\delta_{ \sigma_{1} , \sigma_{2} }
\delta_{ \sigma^{'}_{1}, \sigma^{'}_{2} }
\end{equation}
as it should be.
The six-point function follows similarly.
Thus we have verified all the facts concerning the correspondence with spin.

\end{document}